\documentclass[sigconf, nonacm]{acmart}

\newcommand\vldbdoi{XX.XX/XXX.XX}

\newcommand\vldbavailabilityurl{URL_TO_YOUR_ARTIFACTS}

\usepackage[noend]{algpseudocode}
\usepackage[T1]{fontenc}
\usepackage[utf8]{inputenc}
\usepackage{algorithmicx}
\usepackage{algorithm}
\usepackage{amsmath}
\usepackage{array}
\usepackage{caption}
\usepackage{color, colortbl}
\usepackage{enumitem}
\usepackage{float}
\usepackage{hyperref}
\usepackage{listings}
\usepackage{amsthm}
\usepackage{lstautogobble}
\usepackage{multirow}
\usepackage{nicefrac}
\usepackage{soul}
\usepackage{subcaption}
\usepackage{fancyvrb}
\usepackage{tikz-qtree-compat}
\usepackage{tikz-qtree}
\usepackage{tikz}
\usepackage[frozencache,cachedir=.]{minted}
\usepackage{cellspace}
\usepackage{latexsym}
\usepackage{supertabular}
\usepackage{tablefootnote}
\usepackage{longtable}
\usepackage{wrapfig}
\usepackage{refcount}
\usepackage{xspace}
\usepackage[english]{babel}
\usepackage{alphabeta}
\newtheoremstyle{noIndentDef}  
  {3pt}   
  {3pt}   
  {\normalfont}  
  {}      
  {\bfseries} 
  {.}     
  { }     
  {\thmname{#1}~\thmnumber{#2} \thmnote{[#3]}} 

\theoremstyle{noIndentDef}

\definecolor{blue-cb}{RGB}{0,119,187}
\definecolor{cyan-cb}{RGB}{51,187,238}
\definecolor{teal-cb}{RGB}{0,153,136}
\definecolor{green-cb}{RGB}{34,136,51}
\definecolor{yellow-cb}{RGB}{170,156,49}
\definecolor{orange-cb}{RGB}{238,119,51}
\definecolor{red-cb}{RGB}{238,102,119}
\definecolor{magenta-cb}{RGB}{238,51,119}
\definecolor{purple-cb}{RGB}{170,51,119}
\definecolor{gray-cb}{RGB}{187,187,187}
\definecolor{c_mutable}{RGB}{32,167,178}

\newcolumntype{P}[1]{>{\centering\arraybackslash}p{#1}} 
\newcolumntype{L}[1]{>{\raggedright\let\newline\\\arraybackslash\hspace{0pt}}m{#1}}
\newcolumntype{R}[1]{>{\raggedleft\let\newline\\\arraybackslash\hspace{0pt}}m{#1}}

\newtheorem{definition}{Definition}

\AtBeginDocument{

}

\lstdefinelanguage{sql}{
    sensitive=false,
    keywords={select, from, where, as, in, and, distinct, create, drop, materialized, view, group, by, array_agg,
              left, outer, join, on},
    keywordstyle=\color{green-cb}, 
    morekeywords=[2]{resultdb},
    keywordstyle = {[2]\color{red-cb}},
    morekeywords=[3]{begin, transaction, commit},
    keywordstyle = {[3]\color{purple-cb}},
    string=[b]',
    morestring=[b]",
    stringstyle=\color{orange-cb},
    comment=[l]{--},
    morecomment=[s]{/*}{*/},
    commentstyle=\color{blue-cb},
}

\newfloat{algfloat}{htbp}{lop}
\floatname{algfloat}{Algorithm}

\algrenewcommand{\algorithmiccomment}[1]{\hfill\textcolor{blue-cb}{$\triangleright$ \textit{#1}}}

\AtBeginEnvironment{algorithm}{\small}
\DeclareCaptionFormat{algorithm}{\small\textbf{#1}#2#3}
\begin{document}
\pagestyle{plain}
\title{A Functional Data Model and Query Language is All You Need [Vision]}


\author{Jens Dittrich}
\affiliation{%
  \institution{Saarland University, Germany}
}

\renewcommand{\shortauthors}{Dittrich}

\begin{abstract}
We propose the vision of a functional data model (FDM) and an associated functional query language (FQL). 
Our proposal has far-reaching consequences: we show a path to come up with a modern QL that solves (almost if not) all problems of SQL (NULL-values, impedance mismatch, SQL injection, missing querying capabilities for updates, etc.). FDM and FQL are much more expressive than the relational model and SQL. In addition, in contrast to SQL, FQL integrates smoothly into existing programming languages. 
In our approach both QL and PL  become the `same thing', thus opening up some interesting holistic optimization opportunities between compilers and databases. In FQL, we also do not need to force application developers to switch to unfamiliar programming paradigms (like SQL or datalog): developers can stick with the abstractions provided by their programming language.
\end{abstract}

\maketitle



\section{Introduction}

This paper is inspired by two very recent papers which we believe give a new spin to what databases can and should be.
The first, published at CIDR in January 2025~\cite{Deshpande} (best paper award), is a vision paper that proposes to keep the entity relationship abstraction as a DDL interface to the DBMS.
So rather than \textit{first} translating an entity-relationship model (ERM) to the relational model (RM) to \textit{then} \texttt{CREATE} tables, that paper allows a DBMS to work with the ERM abstraction \textit{directly}.
This has many positive implications including: all semantics of the ERM are preserved.
In addition, we obtain \textit{RM independence}: we can automatically create an optimal RM rather than hand-coding it. 
However, in terms of query language, \cite{Deshpande} suggests to resort to approaches like SQL++\cite{ong_sql++} which however leads to a couple of additional problems including the impedance mismatch with JSON.
A second recent paper (our own previous work), which appeared in June 2025 at SIGMOD~\cite{Nix2025ExtendingSQL}, identifies a long list of problems with SQL which all have the same root cause: SQL forces all result data into a single (possibly denormalized) result relation. 
Therefore we showed how to extend SQL to allow it to return a result \textit{subdatabase}, i.e.~the relations with their subset of tuples from all the input relations of the query that contribute to the result.
Those results are not shoehorned into a single output stream, but are returned as separate streams.
Those two works heavily inspired our paper. Originally, we planned to simply combine those two ideas. But while doing so, we observed that we can come up with a much more versatile data model and query language. Therefore, our paper is more than the  sum of~\cite{Deshpande} and~\cite{Nix2025ExtendingSQL}. It is the product.

This vision paper makes a long list of contributions:
\begin{enumerate}[leftmargin=*]
\item  We present the vision of a functional data model (FDM) and a functional query language (FQL).

\item FDM tears down the boundaries between tuples, relations, and databases when modeling and querying data. Thus, none of these abstractions and the constructs used in a query language operating on FDM are specific for tuples, relations, and/or databases anymore.
All of these abstractions become the same (just on different levels) and hence all of them may be queried and manipulated with the same query language constructs. You can query any relation as if it were a tuple; you can query any database as if it were a tuple; you can query any set of databases as if it were a tuple, a relation, or a database, and so forth.

\item FDM tears down the boundary between data that is stored and data that is computed. 

\item FDM includes features of key, integrity constraints, and indexing as part of its conceptual definition already rather than an afterthought.

\item FQL is an algebra on FDM functions, i.e.~the input to each QL operator is a function and the output is a function. Thus, FQL is not limited to returning a single table as in SQL or relational algebra-inspired languages. 

\item FQL never leaves the data model in order to work around and/or compensate data model representation problems.

\item FQL is as powerful for querying as it is for changing data.  In SQL, querying (reading) the data is much more powerful than writing.
which is limited to modifying sets of tuples of relations or for some DBMS products writing through views (under certain conditions). FQL overcomes these limitations.

\item FQL is easily extensible. Whether a function is defined by `a~user' or by `a library', FQL allows for using functions defined outside the realm of the database. Extending the FQL is as loading a library in Python through an \texttt{import}-statement. 

\item FQL seamlessly blends into host programming languages. Everything in the FQL is expressible through operators.  From the point of view of a programmer, FQL should look like programming constructs of the PL, however, the PL may decide to delegate parts of these constructs to the database system. 

\item FQL makes SQL injection\footnote{The infamous SQL injection attack is still the 3rd most dangerous software weakness~\cite{cwetop25}.} impossible \textit{by design} and not as an afterthought (e.g., through database drivers, prepared statements, and/or input sanitization). 

\end{enumerate}

\section{Overview of Our Approach}

\subsection{Foundations}
We use a simple function definition taken from~\cite{function}:

\begin{definition}[Function] A function $f$ from a set $X$ to a set $Y$ assigns to each element of $X$ exactly one element of $Y$. The set $X$ is called the \textit{domain} of the function and the set $Y$ is called the \textit{codomain} of the function. [wiki] Domains and codomains may be constrained to a type and/or certain conditions.
\label{def:function}
\end{definition}

\begin{definition}[Higher-Order Function] A function $f$ is called a higher-order function if its domain or codomain is a set of functions.
\end{definition}

\subsection{A Functional Data Model (FDM)}

Rather than modeling relations as sets, a central idea of our paper is the following: we \textbf{model everything as a function} --- including tuples, relations, databases, and sets of databases. 
We start with tuples at the lowest level, here the natural input to a function is the attribute name returning the attribute value as the result\footnote{or for pivot tables it may be the individual data values of an attribute of the underlying column; for horizontal partitions of an input relation it could be the partitioning key, e.g.~the attributes specified in GROUP BY.}.
For relation functions, a natural input is either a `row'- or `tuple'-id or their primary key. Calls to relation functions return tuple functions.
For database functions, a natural input is the name of a `table'. Calls to database functions return relation functions. And so forth.

\begin{center}
\begin{footnotesize}
\begin{tabular}{|l||p{3.7cm}|p{1.3cm}|}
\hline
& \multicolumn{2}{c|}{\textbf{Modeling Concept}}  \\ 
\textbf{Abstraction} & \textbf{Relational Model} & \textbf{FDM}  \\\hline\hline
\textbf{tuple} & sequence of attribute/value-pairs, alternatively: function & function \\\hline
\textbf{relation} & set of tuples & function \\\hline
\textbf{database} & set of relations & function \\\hline
\textbf{set of databases} & set of databases & function \\\hline
\end{tabular}
\end{footnotesize}
\end{center}


One main effect of this model is that we can tear down the boundaries between tuples, relations, and databases when modeling, representing, and querying data. Thus, none of these abstractions and the constructs used in a query language operating on FDM are specific for tuples, relations, and/or databases anymore.
All of these abstractions become the same (just on different levels) and hence all of them may be queried and manipulated with the same query language constructs. In other words, you can query any relation as if it were a tuple, you can query any database as if it were a tuple. Likewise you can query any set of databases as if it were a tuple, a relation, or a database, and so forth.

\subsection{Tuple Functions}

In general, a function maps elements from a domain to a codomain (\autoref{def:function}). Thus, a natural conceptual, but discrete, way to define a \textit{tuple function} representing the data of a single tuple is:
\[
t_1(attr:\text{string}) := \big\{ (\text{`name'}:\text{`Alice'}), (\text{`foo'}:12) \big\}.
\]
Function $t_1(X)\mapsto Y$ has the domain $X= \{ \text{`name'}, \text{`foo'} \}$ and the codomain $Y=\{ \text{`Alice'}, 12\}$.
Thus, looking up an attribute value is equivalent to calling a tuple function with the attribute name, e.g.~$t_1(\text{`foo'})= 12$.
This syntax must not be confused with the mapping syntax used in JSON or Python which also use curly brackets to denote `dictionaries' (dictionaries are just one implementation of a map which is in turn a specific  discrete function).

\noindent\textbf{Computed Functions}.  A function does not have to explicitly enumerate the concrete mappings. For instance, we could define a tuple function $t$ to return a computed attribute value: 
\begin{equation*}
t(attr:\text{string}) := \left\{
\begin{aligned}
&42\cdot t_1(\text{`foo'}),& \text{ if } attr=\text{`bar'},\\
&t_1(attr),& \text{ otherwise.}
\end{aligned}\right.
\end{equation*}
This implies, that the value of an attribute can be computed and is indistinguishable from an attribute that is not computed\footnote{This is similar to writing \texttt{SELECT\;42*foo} ... in SQL}.
Technically, we could also go as far as to model every attribute as a function, but we believe that that would overcomplicate our model\footnote{In the 80ies there were a couple of works proposing exactly that. See our discussion in Related Work.}.

In other words, under this functional model the boundary between data that is \textit{stored} and data that is \textit{computed} is removed. 


So, basically, a computed attribute value may return something that was never `inserted' into the database in the sense of a constant value, as `static data' if you wish. 
The idea to model a tuple as a function is in line with database theory~\cite{DBLP:books/cs/Maier83,DBLP:books/aw/AbiteboulHV95}. Note however, that in all these works and also Codd himself~\cite{codd_relational-model}  then proceed to model a relation as a \textbf{set} of tuples. In contrast to that, we proceed to model relations as functions.


\subsection{Relation Functions} 
 
\noindent\textbf{Basic Form.} Let's assume a function $R1$ mapping an input $bar$ to a tuple function $t_{bar}$. Here we assume $bar$ to be a primary key, however, we could also use any other suitable candidate key or a tuple or row-id:
$
R1(bar:\text{int}) := t_{bar}$.
$R1$ is a higher-order function that we call a \textit{relation function}. It represents the data that in the relational model would be kept in a \textit{set} of tuples: a relation. We may call $R1$ with an integer input \textit{bar} and receive a tuple function mapped to from that $bar$ value.
Assume we have two tuple functions:
\[
\begin{aligned}
&t_1(attr:\text{string}) := \big\{ (\text{`name'}:\text{`Alice'}), (\text{`foo'}:12) \big\},\\
&t_3(attr:\text{string}) := \big\{ (\text{`name'}:\text{`Bob'}), (\text{`foo'}:25) \big\}.
\end{aligned}
\]

Now, a call to $R1(1)$ returns $t_1$, a call to $R1(3)$ returns $t_3$. Calls to $bar \notin \{1,3\}$ are not defined.

\noindent\textbf{Constraining Relation Functions}.  
In order to represent which concrete tuples `exist` in relation function $R$, we may constrain the input domain to allow for specific $bar$ values only, e.g.:
$
R(bar:X) \text{ where } X = \{1,3\} \cap \mathbb{N}^+. 
$
These constraints do not have to be expressed as a discrete set of the valid $bar$ values but can alternatively be expressed as a more generic predicate:
$
R(bar:X) \text{ where } X = [7;12] \cap \mathbb{R}^+. 
$
This implies that a relation function may represent a data space that is not just a discrete set but a continuous subspace of `tuple functions', i.e.~the entire codomain spanned under the given domain~$X$.
These constraints can be used to \textit{type} the input and the output of functions and express integrity constraints.

\noindent\textbf{Unique Constraints and Indexes.} The relational model comes with the possibility to add additional constraints like unique constraints. In addition, relational DBMS allow for adding indexes. Both concepts are generalized by relation functions.
Assume we define a second relation function:
$
R2(\text{foo}:\text{int}) := t_{foo}.
$
This provides an alternative view on the tuple functions already returned by $R1(bar:\text{int})$, however organized by attribute $foo$.
The mathematical definition of a function already guarantees that for each $foo$, $R2$ may only point to at most one tuple.
Thus, \textit{the basic function definition of \autoref{def:function} already provides the unique constraint}.
If we want to allow for assigning multiple tuples with a specific input, i.e.~duplicates, we need to explicitly nest the results returned by the relation function, e.g., assume a third tuple function with a duplicate $foo$ value:

$
\begin{aligned}
&t_4(attr:\text{string}) := \big\{ (\text{`name'}:\text{`Thomas'}), (\text{`foo'}:25) \big\}.
\end{aligned}
$

\noindent Then, we need to define $R3(\text{foo}:\text{int}) \mapsto \{ TF \}$. In other words, the relation function $R3$ returns a set of tuple functions.
In a relational DBMS, this is exactly what indexes on attributes with duplicates do! FDM includes this idea on a conceptual level already and not as an afterthought on top of the data model.

\noindent\textbf{Computed Relations.} Let's define a relation function $R4$ as:

\begin{small}
$$
\noindent\begin{aligned}
&R4(bar:\text{int}) :=\\
&\left\{
\begin{aligned}
&t_{bar},& \text{if } bar \in \{1,3\},\\
&  \lambda bar.\big\{ (\text{`name':} \text{rndStr}(seed=bar)), (\text{`foo':} 42\cdot bar) \big\}   ,& \text{ otherwise.}\\
\end{aligned}\right.\\
\end{aligned}
$$
\end{small}

Here, $\lambda$ expresses an anonymous function.
In other words, if a predefined tuple function does not exist in $R4$, i.e., $bar$ is neither 1 nor 3, $R4$ returns an anonymous function: a  $\lambda$-tuple function where the value returned by the name attribute is a pseudo-random (seeded) string.
If `foo' is used as a parameter in that tuple function, that tuple function will return $42\cdot bar$.
For instance, let's retrieve a tuple function from $R4$ and access one of its attributes in a single expression:
$
R4(10)(\text{`foo'}) = 420.
$
In turn, 
$
R4(3)(\text{`foo'}) = 25.
$

Relation function $R4(X)\mapsto Y$ has the domain $X=int$. $Y$ is a set of tuple functions which can be further typed to be restricted to a specific type of tuple function. 

\subsection{Database Functions}
Let's assume a function mapping an input attribute $rel\_name$ to a relation function. For instance,

\begin{small}
\begin{flalign*}
DB(rel\_name:\text{string})  := \big\{ & (\text{`myTab'}:t_4), (\text{`Table1' : } R1), (\text{`Table2' : } R2)  \big\}.
\end{flalign*}
\end{small}

Thus, given the name of a relation, DB returns a relation function. We coin this a \textit{database function}. Just like relation functions, a database function may also return a computed $lambda$ function. In other words, DB may return computed relation functions which were never `stored'.

\subsection{Blurring the lines between the Different Functions} 

Let's extend our example from above. We keep $t_1$ but change $t_3$:
\[
\begin{aligned}
&t_1(attr:\text{string}) := \big\{ (\text{`name'}:\text{`Alice'}), (\text{`foo'}:12) \big\},\\
&t_3(attr:\text{string}) := \big\{ (\text{`name'}:\text{`Bob'}), (\text{`foo' : } t_1) \big\}.\\
\end{aligned}
\]
Now, $t_3$ is a higher-order function. Would we still call $t_3$ `a tuple'?
Definitely not in the sense of the first normal form which mandates that attribute values shall be atomic.
Yet tuple and even relation nesting has been enabled in SQL starting with SQL~99.
Let's extend our example and add a tuple $t_5$ with an attribute returning a relation function when accessing attribute `foo':
\[
\begin{aligned}
&t_5(attr:\text{string}) := \big\{ (\text{`name'}:\text{`Tom'}), (\text{`foo' : } R) \big\}.
\end{aligned}
\]
Tuple $t_5$ semantically feels like adding metadata to a relation. Yet, in the relational model we would still call $t_5$ a tuple and use the modeling constructs available for tuples: `a relation is a set of tuples'. 

In contrast, in our approach, we use the same modeling construct \textit{at all levels}: functions.
Thus, the artificial boundary between tuples, relations, and databases is gone and we could promote $t_5$ to become part of the codomain of a database function.

Whether your `table' $foo$ is an actual relation or: $foo$ is nested in some other table, or $foo$ represents a collection of databases, you can use the same exact abstractions and thus query language constructs.
There is no need anymore to work around this with \texttt{ARRAY}~\cite{Duckdb_array} domains, MAP~\cite{Duckdb_map} domains or other workarounds in your SQL table definitions.

\section{Relationship Functions}

\begin{figure}[t!]
	\includegraphics[trim = 0mm 187mm 400mm 0mm, clip,width=.8\columnwidth,keepaspectratio,page=1]{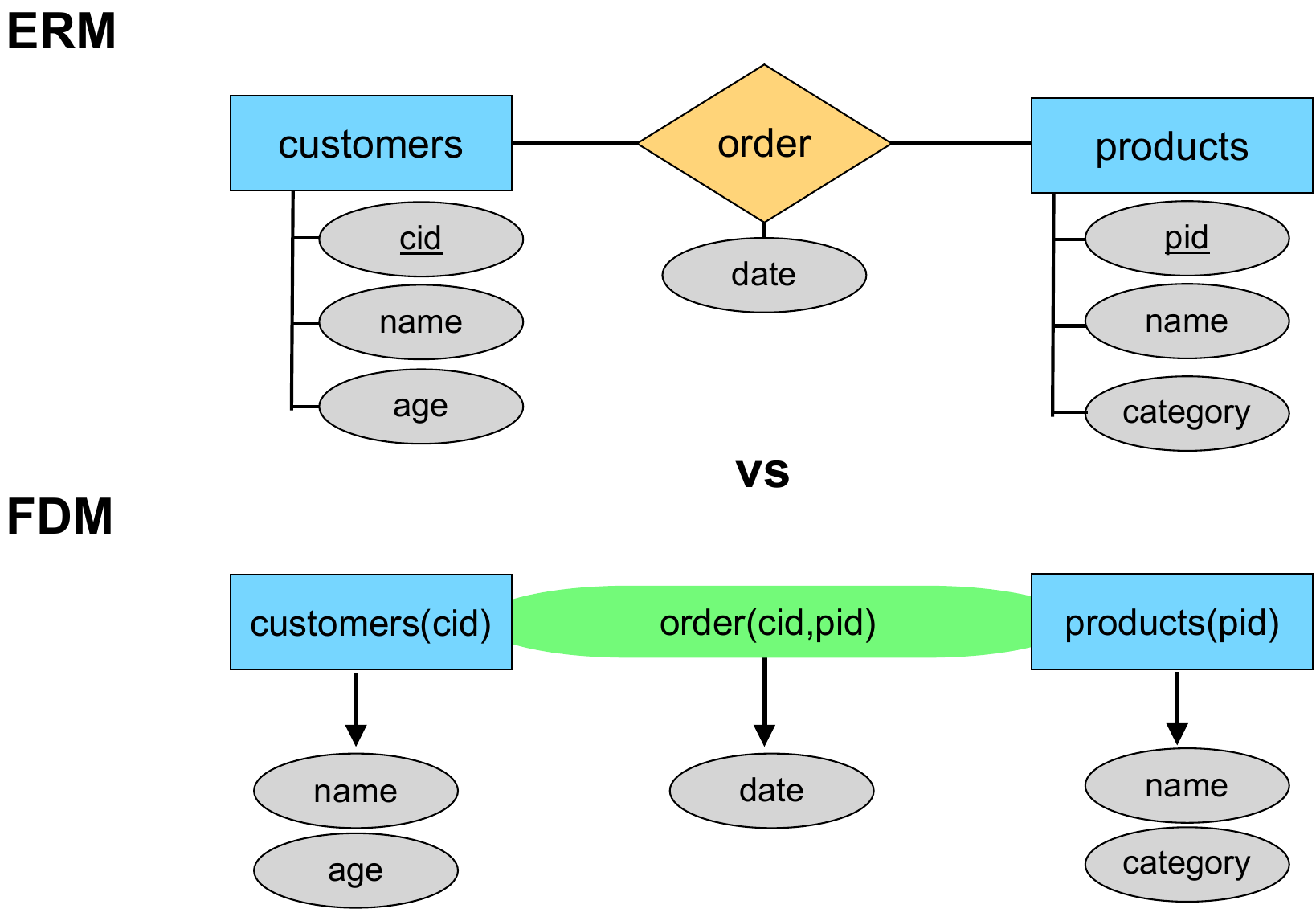}
        \caption{Traditional entity-relationship diagram vs relationship function representation\label{fig:ERvsFDM}}
\end{figure}
In FDM, we can express relationships actually very easily. 

\begin{figure}[t!]
     \centering
	\includegraphics[trim = 0mm 282mm 510mm 0mm, clip,width=.6\columnwidth,keepaspectratio,page=2]{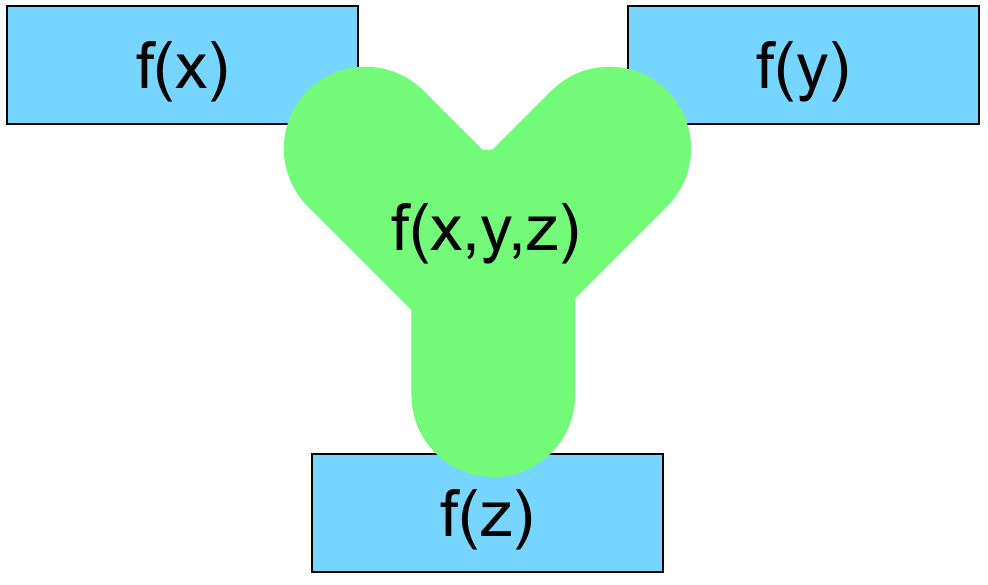}
    \caption{General idea of a Relationship Function\label{fig:FDM:relationship_function}}
\end{figure}

\begin{definition}[Relationship Functions and Predicates]
Given $k$ functions F$_1,\ldots$, F$_k$ with domains $X_1,\ldots,X_k$, a relationship among these functions can be expressed through a relationship function:
\[
\text{RF}(X_1,\ldots,X_k) \mapsto Y_{\text{RF}}.
\]
\label{def:relationship_functon}
If $Y_{\text{RF}}==\texttt{bool}$, we call RF a \textit{relationship predicate} indicating whether a relationship exists among F$_1,\ldots$, F$_k$ for a given input.
\end{definition}
\autoref{fig:FDM:relationship_function} shows the general idea of a relationship function.
\autoref{fig:ERvsFDM} shows an ER-diagram of our running example vs FDM expressing the same semantics.
In FDM, any relationship among functions can be expressed by creating a \textit{relationship function} having as its input the combined inputs of the participating functions.
For our running example, in~\autoref{fig:ERvsFDM}, we have two relation functions customers($cid$) and products($pid$).
To express a relationship among these two relation functions, we can simply define a function order($cid$, $pid$).
Note that the keys $cid$ and $pid$ are not part of the returned attributes.
Also note that we may use any suitable key. In addition, we do not necessarily imply that order($cid$, $pid$) is \textit{stored} as a table as in a relational database.
The latter is a physical design decision, i.e.~FDM may be mapped to a more physical or lower-level data model similar to mapping ERM to the relational model as a separate physiological design step as suggested in~\cite{Deshpande}.

\begin{figure}[t!]
     \centering
	\includegraphics[trim = 0mm 322mm 440mm 0mm, clip,width=.8\columnwidth,keepaspectratio,page=3]{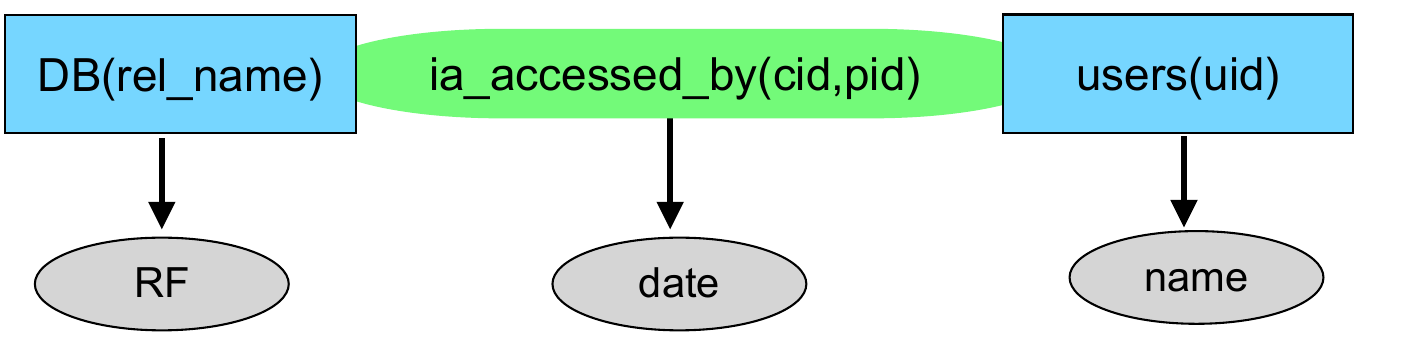}
    \caption{FDM can express a relationship between a database and a relation\label{fig:ERvsFDM:DBrel}, cf.~with ERM which can only model the metadata of a database. }
\end{figure}

In the example,  $cid$ of customers  and $cid$ of order have the same domain. Similarly, $pid$ of order and $pid$ of products have the same domain. This already implies the traditional foreign key constraints we would have to enforce in the relational model/SQL as separate constraints: there, the intermediate `order' relation may only use foreign keys that exist in `customers' and `products'. In FDM, we enforce these constraints as a side effect by simply making functions share the same domains.
Note that \autoref{def:relationship_functon} supports relations between attributes, relations (not only their tuples), databases, etc.
So, does your data model allow you to express a relationship between a database and a relation? Well, FDM can do that: see~\autoref{fig:ERvsFDM:DBrel} for an example. In that figure, we se a database function DB() and a relation function users(). Both are connected through a relationship function is\_accessed(). In other words, this specific FDM expresses which relation (not the relation's metadata!) in the database is accessed by which user, without having to resort to working around this through the database's metadata.

\section{FQL}
\label{sec:MapViews}
 
\subsection{FQL Operator Algebra}
\label{sec:FQL:algebra}

\begin{definition}[FQL Operator]
\label{def:FQL:operator}
The general form of an FQL operator is:
$
Op(\texttt{f}_{\text{in}})\mapsto \texttt{f}_{\text{out}}.
$
\end{definition}
In other words, the input of an operator is a function and the output is a function and thus an FQL operator is simply a higher-order function transforming an input function into an output function.

Note that in sharp contrast to relational algebra operators (e.g.~ $\pi$, $\sigma$, $\cup$, $\times$, $\bowtie$, ...), the input data \textbf{and} the output data of an FQL operator \textbf{is a function}. 
Thus, FQL operators are neither restricted to representing a single output relation nor restricted to up to two input relations as in relational algebra. FQL operators can represent any granule including tuples, relations, databases, sets of databases, etc.
Our approach also goes beyond our own recently proposed \texttt{RESULTDB} extension~\cite{Nix2025ExtendingSQL} in that we cannot only return a relational subdatabase but can also return a database with a completely different schema, e.g.~for aggregations.

\begin{definition}[FQL Algebra]
\label{def:FQL:algebra}
Our FQL algebra consists of
\begin{enumerate}
\item FDM functions (as operands)
\item FQL functions (as operators)
\item rules for combining them (nested operators).
\end{enumerate}
This algebra defines how to assemble operators into \textit{FQL expressions}.
\end{definition}

We do not imply that an FQL expression is executed in the order specified. 
In relational algebra there is often the confusion whether a relational algebra expression is supposed to be read as a declaration (not implying any order) or a function (implying a specific order).
In FQL we allow for both interpretations and if in doubt mark which interpretation is meant.
From the point of view of a programmer an FQL expression simply looks like a nested function call.

\subsection{FQL Operator Costumes}
\label{sec:FQL:operatorcostumes}

FQL does not impose any new syntax on programmers. We envision FQL as a `bunch of functions' that use the type system of the embedding programming language. So the same FQL expression may look syntactically different in Python and C++ or Rust. Yet, conceptually that PL syntax can express the same query semantics. We call an FQL operator in a host PL a \textit{function costume}. These functions are \textbf{not necessarily} compiled to machine code or interpreted by the PL's compiler and/or runtime. Instead the entire FQL expression or any suitable part of it \textbf{may be pushed down} to the database system which can then optimize the expression and return a function (through some streaming interface: ONC, generators, vectorized, etc.).
So from the point of view of the developer it looks like as if all those functions were executed in the PL --- and in the order specified. However, the underlying team of PL compiler/runtime and DBMS may decide differently.
Therefore, FQL avoids the classical impedance mismatch between query languages and programming languages (PL) types where the embedded SQL feels alien to the PL. In fact, the entire artificial boundary between SQL and the PL embedding SQL-expressions vanishes.
This also opens up interesting future work as now the optimizations done by the MyPL compiler/interpreter/runtime overlap with the optimizations done with the DBMS and may be done in the same optimization space.
Thus, we now have the option to delegate parts of the query expression down to the DBMS and leave some to the PL, e.g.~depending on runtime statistics or the likelihood that DB optimizations may even have a benefit. In addition, note that this implies that the \textbf{artificial boundary between an embedded PL-statement (a~UDF) and the outer query is gone}, too.
This opens up exciting future work on how to integrate these optimization and execution steps.

\subsection{FQL Examples}

Figures~\ref{fig:Python:grouping}--\ref{fig:Python:updates}. show example of FQL operator costumes reflecting FQL expressions in Python.
We assume a simple schema of customers ordering products as shown in~\autoref{fig:ERvsFDM}.
\autoref{fig:Python:filter} shows different alternatives for filtering data. \autoref{fig:Python:grouping:single_steps}\&\ref{fig:Python:grouping_partially_fused} show how to express grouping and aggregation.
\autoref{fig:Python:subdatabase} shows how to declare a subdatabase. This is the FQL version of the SQL-extension proposed in~\cite{Nix2025ExtendingSQL}. \autoref{fig:Python:subdatabase_with_on} shows a classical join producing a single-table result.
\autoref{fig:Python:subdatabase_outer} shows a generalized outer join semantic for subdatabases. \autoref{fig:Python:subdatabase:grouping_sets} 
shows how FQL keeps the semantically different subresults of rollup and cube in separate relation functions,
\autoref{fig:Python:DB_set_operations} illustrates how to perform set operations on entire databases rather than just relations.
\autoref{fig:Python:updates} and \ref{fig:Python:transactions} show how to express change operations and transactions.

\begin{figure}[th!]
\begin{footnotesize}
 \begin{subfigure}[b]{\columnwidth}
\begin{minted}{python}
# customers is a relation function obtained from database function DB:
customers: RelationF = DB('customers')                                           
# alternatively:
customers: RelationF = DB.customers 
# get customers older than 42 years, function syntax:
customers_42: RelationF = filter(lambda prof: prof("age") > 42, customers)            
# alternatively, dot syntax:
customers_42: RelationF = filter(lambda prof: prof.age > 42, customers)            
# alternatively, Django ORM-style:
customers_42: RelationF = filter(age__gt=42, customers)            
# alternatively, break up the predicate:
from operators import *
customers_42: RelationF = filter(att='age', op=gt, c=42, customers)            
# alternatively, textual predicate with free parameters:
customers_42: RelationF = filter("age>$foo", {foo: 42}, customers)            
\end{minted}   
\caption{six different syntax alternatives to declare all customers older than 42 years}
\label{fig:Python:filter}
\end{subfigure}

 \begin{subfigure}[b]{\columnwidth}
\begin{minted}{python}
# define a DB of relation functions representing age_groups:
groups: DBF = group(lambda prof: prof.age, customers)
# alternatively, using an explicit 'by' parameter:
groups: DBF = group(by=["age"], customers)
# compute one aggregate per input group and
# declare new attributes for the output:
aggregates: RelationF = aggregate(count=Count(), groups)
#  get groups with more than 9 entries:
large_groups: RelationF = filter(lambda g: g.count > 9, aggregates)     
\end{minted}   
\caption{Option 1: grouping, aggregating, and having unrolled as individual steps}
\label{fig:Python:grouping:single_steps}
\end{subfigure}

\begin{subfigure}[b]{\columnwidth}
\begin{minted}{python}
aggregated_ages: DBF = group_and_aggregate(                                  
                          by=["age"],  count=Count(), input=customers)
large_groups: RelationF = filter(lambda g: g.age > 9, aggregated_ages)     
\end{minted}   
\caption{Option 2: grouping plus aggregation as a single step (corresponds to GROUP BY syntax in SQL)}
\label{fig:Python:grouping_partially_fused}
\end{subfigure}
\end{footnotesize}
\caption{\label{fig:Python:grouping}Python FQL operator costumes  alternatives, including Python typehints}
\end{figure}

\begin{figure}[th!]
\begin{footnotesize}
\begin{minted}{python}
# pick a subset of relations for the subdatabase:
relations: list[string] =  ['order', 'products']
subdatabase: DBF = filter(lambda kv: kv[0] in relations, DB)      
# add customers_NY to subdatabase:
subdatabase.customers = filter(state='NY', DB.customers)
# reduce the database:
subdatabase_reduced: DBF = reduce_DB(subdatabase)
\end{minted}   
\end{footnotesize}
\caption{\label{fig:Python:subdatabase}ResultDB Query producing a subdatabase}
\end{figure}

\begin{figure}[th!]
\begin{footnotesize}
\begin{minted}{python}
# join the database along the foreign key constraints in the schema:
# the join may be performed in any suitable way chosen by the optimizer
# including n-ary joins:
join_result: RF = join(subdatabase)
# alternatively, specify the join conditions explicitly
join_result: RF = join(subdatabase, 
                      on=[  [customers.id, order.c_id],
                            [order.p_id, products.id]]
                  )
\end{minted}   
\end{footnotesize}
\caption{\label{fig:Python:subdatabase_with_on}Join a subdatabase of n relations returning a single denormalized relation function}
\end{figure}

\begin{figure}[th!]
\begin{footnotesize}
\begin{minted}{python}
# mark the relation(s) to return outer semantics:
subDB: DBF = subdatabase(outer='products', subdatabase)
# obtain references to the two semantically different relations:
products_unsold: RF = subDB.products.outer
products_sold: RF = subDB.products.inner
\end{minted}   
\end{footnotesize}
\caption{\label{fig:Python:subdatabase_outer}Outer join separating outer from inner semantics in the output; cf.~outer joins in SQL forcing the result into a single output relation. Note that the terms `left' and `right' outer join do not make sense here, as we are not restricted to binary joins like in SQL. Instead, we mark which relations are to be returned with both inner and outer semantics. }
\end{figure}

\begin{figure}[th!]
\begin{footnotesize}
    \begin{minted}{python}
gset: DBF = group_and_aggregate(  
        {(by:["age"], count:Count(),          name: "age_cc"),
         (by:["age","name"], count:Count(),   name: "age_name_cc"),
         (by:[], min:Min(),                   name: "global_min")
        }, input=customers
    )
# obtain three references to the semantically different groupings:
RFx, RFy, RFz:= gset.age_c, gset.age_name_c, gset.global_min
\end{minted}   
\end{footnotesize}
\caption{\label{fig:Python:subdatabase:grouping_sets}Grouping sets outputting different relations per semantically different grouping condition; cf.~grouping sets in SQL forcing the result into a single output relation and thus filling up the result with NULL-values. Examples for the cube and rollup operator are similar.}
\end{figure}

\begin{figure}[th!]
\begin{footnotesize}
\begin{minted}{python}
# create a deep copy of the database DB:
DB_copy: DBF = deep_copy(DB)
# now change DB_copy by inserting, adding tuples, tables, etc.
...
# compute the differential database just showing changes
DB_diff: DBF = difference(DB, DB_copy)
# similar for intersect:
DB_intersect: DBF = intersect(DB, DB_copy)
# similar for minus and union:
DB_minus, DB_union = minus(DB, DB_copy), union(DB, DB_copy)
\end{minted}   
\end{footnotesize}
\caption{\label{fig:Python:DB_set_operations}Compute set operations on multiple input databases rather than on individual relations}
\end{figure}

\begin{figure}[th!]
\begin{footnotesize}
\begin{minted}{python}
# get references to relation function 'customers':
customers: RelationF = DB.customers
# adding a 'tuple', i.e. a tuple function:
customers[3] = {'name':'Tom', 'age': 42}
# alternatively, insert relying on an auto id:
customers.add( {'name':'Stephen', 'age': 28} )
# updating a 'tuple':
customers[3] = {'name':'Tom', 'age': 49}
# updating a an attribute value of a tuple:
customers[3]['age'] = 50
# delete a tuple function from relation function 'customers':
del customers[3]
\end{minted}   
\end{footnotesize}
\caption{\label{fig:Python:updates}inserts, updates, and deletes; note the absence of an explicit save()-method: changes are applied immediately to the snapshot: depending on the configured transaction mode either the snapshot of the transaction, or: the snapshot of the individual operation (statement)}
\end{figure}

\begin{figure}[th!]
\begin{footnotesize}
\begin{minted}{python}
begin()
# get references to relation function 'accounts':
accounts: RelationF = DB.accounts
# remove 100 Euros from account 42
accounts[42]['balance'] -= 100
# add 100 Euros to account 84
accounts[84]['balance'] += 100
commit()
\end{minted}   
\end{footnotesize}
\caption{\label{fig:Python:transactions}Transactional snapshot semantics in Python using a bank accounts example
 }
\end{figure}


%
%

\subsection{In-Place FQL Usage}

\label{sec:FQL:inplace}

Any FQL expression can be used in two different ways:

\noindent\textbf{(1.)~Out-of-place Usage:} The FQL expression offers a different \textit{perspective} in the sense of a database \textit{view} on the input. It does not change the input function in the underlying relation or database, but based on that input function, the FQL operator returns an output function conceptually reflecting a consistent snapshot.  This corresponds to a classical read-only (\texttt{SELECT}) query copying data from the database to the outside by means of producing a result set. This leaves the data in the database unchanged. Keep in mind that any  SQL-\texttt{SELECT} query is already simply a one time view on the data, too.

\noindent\textbf{(2.)~In-place Usage:}  This FQL expression replaces a function in the input FDM. In this mode, the FQL expression is used as a ``data rewrite rule''. Such a rule may mimic SQL's database DML-operations like \texttt{INSERT}/\texttt{UPDATE}/\texttt{DELETE}. However, in SQL, the latter operations are limited to modifying one or multiple tuples of a relation. In our approach, we can transform and replace the underlying functions arbitrarily. Conceptually, we can replace entire relations or even databases with a single expression. With in-place FQL expressions the full expressive power of FQL can be leveraged to transform the underlying data. 

\noindent\textbf{Example.}  Let $MyRel$ be a relation function. Then,
$
DB(\text{`otherRel'}) :=  MyRel
$
can be used to add that relation function to database function $DB$ (replacing any other previous mappings in $DB$ from \text{`otherRel'} on the way). The same holds for any FQL expression \textit{foo}, no matter how complex \textit{foo} may be, you may assign it to $DB$ as follows:
$
DB(\text{`myAwesomeView'}) :=  foo.
$
Similar to relational databases and the distinction between dynamic and materialized views, we need to decide though whether to make these assignments dynamic or whether we materialize their contents. We assume that all those assignments are dynamic unless explicitly marked with a copy-function:
$
DB(\text{`myAwesomeMaterializedView'}) :=  \text{copy}(foo).
$
This will materialize the contents of $foo$. It is equivalent to a deep copy-operation and comes with all the trade-offs known for traditional materialized views (storage requirements, maintenance, freshness).


\section{Related Work}
\label{sec:RW}

\noindent\textbf{Beyond Tables and Single Table Results.} See our discussion of~\cite{Deshpande} and ~\cite{Nix2025ExtendingSQL} in the Introduction. Note that the idea of returning a subdatabase was also explored from an information retrieval viewpoint outside SQL in~\cite{10.1007/s00778-007-0075-9,10.1145/1353343.1353396}.

\noindent\textbf{Query Language Alternatives.} In terms of relational QLs there were other proposals that basically boil down to proposing a hybrid of RA operators and SQL-style syntax in the PL~\cite{piglatin,linq,PRQL,ShuteBBBDKLMMSWWY24}; a variant of RA  in the PL, e.g.~XXL~\cite{10.5555/645927.672371}; or offer both RA and SQL~\cite{ZahariaCDDMMFSS12}, and/or combine that with a pipe syntax~\cite{PRQL,ShuteBBBDKLMMSWWY24}.
The most recent call to replace SQL with something more functional is~\cite{NL24}.
However, that work builds heavily on relational algebra and allows developers to build pipelines on the relational model through a dot syntax, very similar to Django ORM QuerySets which always starts with one relation and the appends additional relations through joins.
Such a syntax is too limited for our data model as we want our operators to be able to express transformations on entire (arbitrary) `subdatabases' (any nested function expression) rather than just single relations.
There were a couple of other approaches trying to break the RA operator abstraction into suboperators~\cite{Lohman88}, \cite{DittrichN20}, and \cite{BandleG21,KohnL021}. 
In fact, any FQL operator doing `less' than producing a single output relation can be coined a \textbf{suboperator}. Similarly, any FQL operator returning something bigger than a relation could be called a \textbf{superoperator}.

\noindent\textbf{Previous Functional Data Models and Languages.} In the 1970ies and 1980ies, there has been quite a number of approaches on developing functional data models and languages.
See Peter M.~Gray's  excellent encyclopedia articles~\cite{DBLP:reference/db/Gray18,DBLP:reference/db/Gray18c,DBLP:reference/db/Gray18d} and book~\cite{DBLP:journals/program/Gray05} for a start. Though the terms FDM~\cite{DBLP:conf/sigmod/Shipman79,DBLP:journals/cj/KulkarniA86} and FQL~\cite{DBLP:conf/sigmod/BunemanF79} were the same as the ones we use, FDM back then was more of a triplet data model and RDF ``[was] very similar to the Functional Data Model''~\cite{DBLP:reference/db/Gray18c}. Though the principle functional view of data back then was similar to what we are proposing now; back then, most of the query languages proposed were still textual and considered separate from the programming language, e.g.~OSQL~\cite{Beech88}. An exception was Adaplex~\cite{CDF87} which integrated a query language into Ada, however made use of nested loop syntax to express query semantics rather than more declarative operators like in our FQL. Many of these works can be seen as predecessors to which then became for object-oriented DBMS,  object-relational DBMSes, and ORMs~\cite{10.1145/92755.92772} as workarounds for the mismatch of the relational and the object-oriented world.

\noindent\textbf{Mixed model approaches.} Another line of work tries to marry the relational world with JSON, e.g.~SQL++\cite{ong_sql++} and Oracle~\cite{oracle,10.1145/3722212.3724441}.
However, as outlined by~\cite{Nix2025ExtendingSQL}, these approaches cannot represent N:M result sets. All those approaches mix up the underlying conceptual (mathematical) data model (tree structured data) with its representation (the syntax).
This common confusion was already discussed in other proposals for unifying data models like~\cite{DS06}.
Rel~\cite{rel,Rel25} has a very similar motivation as our work. However, in contrast to FDM, Rel keeps the abstraction of relations to be \textit{sets of tuples} and tuples to be \textit{``an ordered immutable sequence of data values''}~\cite{rel} which we both discard to be functions. In addition, in Rel it is unclear how attributes are typed. In contrast, our approach can directly leverage the typing mechanisms of the embedding PL, e.g.~the type hint system in Python wich can even be checked at runtime~\cite{typeguard}.
Syntax-wise, Rel is quite different from the most common PLs like Rust, C++, Python, etc.~and thus is much harder to integrate into those PLs. 
In the 1980ies, a number of interesting proposals were made to develop query languages directly on ERM~\cite{ParentS84,ElmasriW81,CampbellEC85,CzejdoERE90}. Yet all of these works fall behind our approach in terms of expressiveness.

\noindent\textbf{Criticism of SQL, RM, ERM, and RA.}  Since the dawn of the relational model~\cite{codd_relational-model}, SQL~\cite{ChamberlinB74}, and the entity-relationship model~\cite{Chen75}, there has been criticism, e.g.~\cite{date_critique-sql,codd_normalization,NL24,ShuteBBBDKLMMSWWY24}.
Also note our predecessor report~\cite{dittrich2025ridsqlrelationalalgebra} of this paper which identifies and summarizes many of the existing problems with our database foundations. There, we also proposed the RMTM data model based on `maps'. In retrospect, RMTM is somewhat unnecessarily complicated.
In addition, RMTM can easily be confused with its various realizations (thanks to the anonymous SIGMOD 2026 reviewers for pointing this out). Therefore, we developed this second version (keeping in spirit the same ideas) solely using functions and yielding FDM and FQL and thus drawing a clear line between its mathematical formulation and its realizations.

\section{Conclusion}

This paper presented a vision for a functional data model and a functional query language. We believe that our proposal --- though challenging the successful status quo of the relational model, relational algebra, SQL, and ORMs --- comes with a lot of opportunity to see data management from a different viewpoint.
A major feature of our data model (FDM) is that we use the same abstractions at all levels. This opens the book for a query language (FQL) that supports operators unheard of in SQL.

A lot of exciting work lies ahead: including (1.)~a full-blown implementation of FQL and integration into programming languages like Python. (2.)~exploring the joint optimization space, (3.)~extending the list of FQL operators that allow functionality beyond SQL.

\bibliographystyle{ACM-Reference-Format}
\bibliography{main}

\end{document}